# Exploration of Surgeons' Natural Skills for Robotic Catheterization

Olatunji Mumini Omisore, Wenjing Du, Tao Zhou, Shipeng Han, Kamen Ivanov, Yousef Al-Handarish, and Lei Wang

*Abstract*— Despite having the robotic catheter systems which have recently emerged as safe way of performing cardiovascular interventions, a number of important challenges are yet to be investigated. One of them is exploration of surgeons' natural skills during vascular catheterization with robotic systems. In this study, surgeons' natural hand motions were investigated for identification of four basic movements used for intravascular catheterization. Controlled experiment was setup to acquire surface electromyography (sEMG) signals from six muscles that are innervated when a subject with catheterization skills made the four movements in open settings. k-means and k-NN models were implemented over average EMG and root means square features to uniquely identify the movements. The result shows great potentials of sEMG analysis towards designing intelligent cyborg control for safe and efficient robotic catheterization.

*Keywords*: Robotics, Catheterization, sEMG, Signal processing

## I. Introduction

Cardiovascular diseases (CVDs), defined as mal-conditions of the heart or its vessels, remain the leading cause of death and disability. Nearly 31% of annual global deaths are due to CVDs in the last decades [1]. Minimally invasive surgery was adopted for cardiovascular interventions by steering endovascular tools, such as catheters and guidewires, via single port made on the peripheral vessel into the cardiac cavity. However, this technique is embroiled with several challenges such as radiation exposure and spine injuries suffered by both patients and surgeons. Thus, some robotic platforms including Niobe® Magnetic Navigation System (Stereotaxis Inc. USA), Amigo® Robotic Catheter System (RCS; Catheter Precision, USA), Sensei® X (Hansen Medicals and CorPath robotic systems (Corindus Vascular Robotics, USA); were recently introduced into the Cath Labs. The RCSs have achieved reduction in the operational hazards suffered during cardiovascular interventions; however, achieving the natural manipulations skills are still lacking [2].

Recently, application of signal-based control is found applicable in robotics control. Some studies were reported by Zhou *et al.* [3]; however, those studies were majorly focused on robotic systems that have compliant joint structures. These mechanisms are quite different from cases of robotic catheter systems (RCS) used for vascular surgery. Furthermore, most RCSs are operated in master-slave conventions. Thus, to compensate for the novel patient-side slave designs, it is important to manipulate endovascular tools closely to how surgeons manipulate their hands for manual catheterization. A number of ergonomic master interfaces have been proposed to replicate the natural motion patterns of surgeons [4]; however, catheter navigation along tortuous vessels is hard to achieve.

Studies in the field of signal processing have established that hand movements can be characterized with activities of the coordinating muscles. Thus, surface EMG (sEMG) signals have been used for technical skills assessment in intravascular interventions [3]. In a similar vein, the motion patterns and natural skills adopted by surgeons to manipulate endovascular tools (such as guidewire or catheter) can be studied and adapted for efficient robotic catheterization when performing cardiac interventions. Thus, this study is tailored for analysis of muscles activities to identify the patterns associated with hand movements of surgeons during vascular catheterization.

Remainder of this paper is organized such that. Materials and methodology for sEMG signal acquisition are presented in Section II. Movement identification with sEMG modeling and results are discussed in Section III. Finally, conclusion of the study and future directions are highlighted in Section IV.

## II. System Design and Data Acquisition Experiment

Recently, several prototypes of two degree-of-freedom (2-DoF) interventional robots have been designed to mimic hand motions of surgeons when performing navigation actions for vascular catheterization. CAD model of the second generation robot designed for vascular interventions in our lab is shown in Fig. 1. The RCS is smarter and light-weighted than the first generation [2,5]. To pioneer intelligent cyborg control for safe and efficient catheterization by robotic systems, identification of the distinct hand movements made by surgeon is observed by pre-processing effective signals acquired from six muscles of interest, and extraction of useful and reliable features.

### A. Experimental setup and Signal Acquisition

In this study, a controlled experiment was organized to acquire sEMG signals from one subject who is right-handed, without neuromuscular diseases, and have studied about vascular catheterization in the last two years. The age and body mass index of the subject are 33 years and 25.5 kg/m$^2$, respectively. The subject was made to learn the procedure and gave a signed informed consent form (SIAT-IRB-190215-H0291) approved by the Institutional Review Board of Shenzhen Institutes of Advanced Technology, CAS, China.

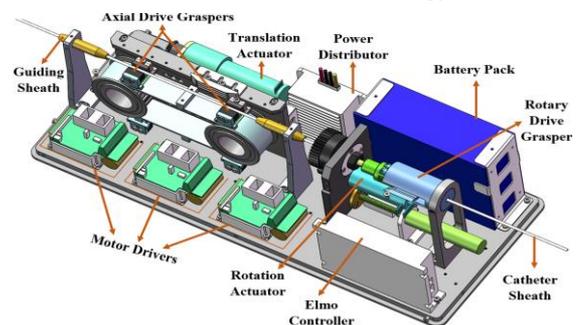

Fig. 1: CAD model of robotic system for cardiovascular intervention.

This work was supported by National Natural Science Foundation of China under Grants #U1505251 and #U1713219; the Key Research Program of the CAS (Grant #KFZO-SW-202); the National High-Tech R&D (863) Program (Grant #2015AA020933); and the outstanding Youth Innovation Research Fund of SIAT-CAS (Grant# Y8G0381001).

W.J. Du, O.M. Omisore, T. Zhou, S.P. Han, K. Ivanov, Y. Al-Handarish and L. Wang are with Research Center for Medical Robotics and MIS Devices, Shenzhen Institutes of Advanced Technology, CAS, Shenzhen, China; { omisore, wj.du, sp.han, tao.zhou, kamen, yousef, wang.lei}@siat.ac.cn.

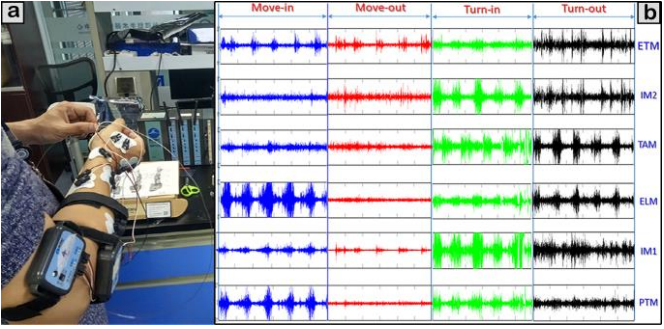

Fig.2: Views from signal acquisition: a) Experimental setup; b) signal plots

A commercial configurable electromyography (EMG) system (BioNomadix, BIOPAC Systems, Inc., Goleta, CA, USA), integrated with the ACQKnowledge software (version 4.2.0) was used for signal acquisition. The data setup consists of 6-channels with disposable Al-AgCl electrodes distributed on the six muscles, as shown in Fig. 2a. To ensure high signal quality, skin of the subject was properly cleaned with alcohol prior to experiment. During experiments, subject was asked to make the four basic movements namely: *move-out*, *move-in*, *turn-in*, and *turn-out* in open setting so as to acquire his signal. Then, the EMG signals were acquired at 1000 Hz sampling rate using a pre-designed band pass filtering with range of 10 to 500 Hz, and 50 Hz notch filter to remove power-line noise from the EMG signals. Each movement was done 100 times and made in two sequences. Each sequence was captured as sEMG signals from the six muscles and saved into separate files for further processing. The sEMG activity recorded in the subject, during the four movements, is plotted in Fig. 2b.

### B. Signal Processing and Feature Extraction

Aside the online processing done during signal acquisition, the sEMG signals were further processed in Matlab 2014 for extraction of reliable features needed for the identification task. These include: (1) segmentation of the signal into a series of 100 ms window length and increment of 50 ms for spectrum analysis with fast fourier transform of each data window used for outlier removal; and (2) normalization of the signal using maximum value normalization approach [6] to establish a baseline for comparing parts of the signal from the subject.

Sequel to signal processing and normalization, Average EMG (AEMG) and root Means Square (RMS) features were extracted from the signals, using Eq. 1 and 2, respectively. AEMG value is an important indicator for a selected length of the EMG signal with data points (N) since it does not have relationship with the length of time when comparing two signals. RMS was considered to provide insight on the amplitude of the sEMG signals since it gives a measure of the power of the signal in a certain time period (T).

$$AEMG = \frac{\sum_{i=1}^{N}|data[i]|}{N} \quad (1)$$

$$RMS = \sqrt{\frac{\sum_{i=1}^{T} data[i]^2}{T}} \quad (2)$$

### III. MOVEMENT IDENTIFICATION WITH SEMG MODELING

To identify the unique movements in the EMG signals that were acquired from the subject, algorithmic models for clustering and classification of the signals were implemented.

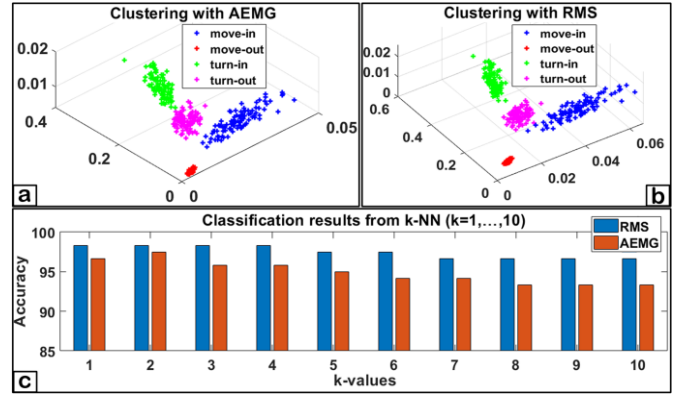

Fig.3: Movement identification results from signal acquisition using:
a-b) k-means (k=4) clustering with AEMG and RMS; c) k-NN classification

This includes k-means clustering and k-nearest neighbour (k-NN) classification models, which were implemented in Matlab 2014 to distribute the data points in the signals into distinct groups. The clustering results observed with k-means (k = 4), when both AEGM and RMS features are used for the subject, are presented as shown in Figs. 3a-b. Similarly, k-NN classification was applied to classify each of the movement by estimating the local density in nearest neighborhoods of each cluster. Values of k were varied between 1 and 10 and the results obtained are plotted as shown in Fig. 3c.

Applicability of sEMG towards the robotic control for endovascular movements can be concluded viable with results of the algorithmic k-means and k-NN models presented in the plots of Fig. 3. By visual inspection, it can be seen that each movements are uniquely identified in distinct clusters. Also, analysis from results shows that both AEMG and RMS features are highly comparable in terms of density ratio and mean amplitude for all movements across the data points, but RMS achieve a better within-cluster distance. Similarly, Fig. 3c shows that high classification results were obtained, and the best scenario was observed when k=2 in the k-NN algorithm.

### IV. CONCLUSION

This pilot study was carried out to verify if sEMG signals can be used to explore surgeons' natural skills for robotic catheterization. For this purpose, muscle activities of a trained subject were correlated with distinct hands movements used for vascular catheterization. Clustering and classification methods were implemented to uniquely identify each data points. Our objective is to transmit the signals to control the 2-DoF RCS for vascular intervention in the future.

REFERENCES

[1] E. Benjamin, M. Blaha, S. Chiuve, M. Cushman, and S. Das, *et al.*, *Heart Diseases and Stroke Statistics*, Circulation, 135(10):e446, 2017.
[2] O. Omisore *et al.*, *Towards Characterization and Adaptive Compensation of Backlash in a Novel Robotic Catheter System for Cardiovascular Interventions*, IEEE TBioCAS, 12(4):824-838, 2018.
[3] X. Zhou, G. Bian, X. Xie, Z. Hou, "*An Interventionalist-Behavior-Based Data Fusion Framework for Guidewire Tracking in Percutaneous Coronary Intervention*", IEEE Trans on SMC: Syst. 2018(PP):1-14.
[4] Raffi-Tari *et al.*, *Objective Assessment of Endovascular Navigation Skills with Force Sensing*, Annals of Biomed Eng, 45(5):1315-27, 2017.
[5] O. Omisore, S. Han, L. Ren, and L. Wang, *A Teleoperated Robotic Catheter System with Motion and Force Feedback for Vascular Surgery*, 18th IEEE ICCAS, South Korea, October 17-20, 2018.
[6] W. Du, O. Omisore, H. Li, K. Ivanov, S. Han, and L. Wang, *Recognition of Chronic Low Back Pain during Lumbar Spine Movements Based on Surface Electromyography Signals*, IEEE Access, 2018(6):65027-42